\documentclass[twocolumn,prb]{revtex4}
\usepackage{amsmath}
\usepackage{amsfonts}
\usepackage{amssymb}
\usepackage{graphics}
\usepackage{float}
\usepackage{graphicx}
\usepackage{epstopdf}
\usepackage[compact]{titlesec}
\usepackage{natbib}
\usepackage{threeparttable}
\usepackage{lipsum}
\usepackage[titletoc,title]{appendix}
\usepackage{titlesec}

\begin{document}
\title{Numerical guidelines for setting up a general purpose k.p simulator with applications to quantum dot heterostructures and topological insulators}
\author{Parijat Sengupta$^{1}$}
\email{psengupta@cae.wisc.edu}
\author{Hoon Ryu$^{2}$}
\author{Sunhee Lee$^{3}$}
\author{Yaohua Tan$^{4}$}
\author{Gerhard Klimeck$^{4}$}
\affiliation{$^{1}$Dept of Material Science and Engineering, University of Wisconsin, Madison, WI 53706 \\
$^{2}$National Institute of Supercomputing and Networking, Korea Institute of Science and Technology Information, Daejeon 305-806, Republic of Korea.\\
$^{3}$Samsung Advanced Institute of Technology, Yongin, Gyeonggi-do 446-712, Republic of Korea.\\
$^{4}$Dept of Electrical and Computer Engineering, Purdue University, 
West Lafayette, IN, 47907 
} 

\begin{abstract}
The k.p perturbation method for determination of electronic structure first pioneered by Kohn and Luttinger continues to provide valuable insight to several band structure features. This method has been adopted to heterostructures confined up to three directions. In this paper, numerical details of setting up a k.p Hamiltonian using the finite difference approximation for such confined nanostructures is explicitly demonstrated. Nanostructures belonging to two symmetry classes namely cubic zincblende and rhombohedral crystals are considered. Rhombohedral crystals, of late, have gained prominence as candidates for the recently discovered topological insulator (TI) class of materials. Lastly the incorporation of strain field to the k.p Hamiltonian and matrix equations for computing the intrinsic and externally applied strain in heterostructures within a continuum approximation is shown. Two applications are considered 1)Computation of the eigen states of a multi-million zincblende InAs quantum dot with a stress-reducing InGaAs layer of varying Indium composition embedded in a GaAs matrix and 2)Dispersion of a rhombohedral topological insulator Bi$_{2}$Se$_{3}$ film. 
\end{abstract}
\maketitle

\section{Introduction}
\label{intro} 
The physics of semiconductors is directly influenced by the carriers in the extrema of various energy bands such that only the neighbourhood of the band extrema are important. The k.p perturbative~\cite{voon2009kp,kane1966k,tsidilkovskii1982band} method and the theory of invariants~\cite{bir1974symmetry} allow us to compute the band structure and shape of energy surfaces of semiconductors from symmetry arguments. The goal of this paper is not to describe the analytical construction of a bulk k.p Hamiltonian but demonstrate numerical techniques to adapt the Hamiltonian to a nanostructure~\cite{bastard1988wave,harrison2005quantum} and obtain relevant eigen values and wave functions. 

The paper is structured as follows: In Section ~\ref{II}, the commonly used bulk eight-band k.p Hamiltonian for direct band gap zincblende semiconductors is introduced. This Hamiltonian accurately describes the eigen states around the $ \Gamma $ point in \textit{k}-space. A truncated four-band Hamiltonian with a limited basis set is also outlined to study dispersion relationship in the newly discovered topological insulators(TI). A detailed description of the finite-difference discretization approach as applied to k.p Hamiltonians is given in Section.~\ref{III}. A matrix equation for computing electronic strain within the continuum method framework is established in Section.~\ref{IV}. In Section ~\ref{V}, discretized k.p Hamiltonians are employed to estimate the eigen states of a zincblende quantum dot heterostructure and topological insulator thin-films. The paper concludes by briefly evaluating the benefits and drawback of a perturbative k.p approach over atomistic tight-binding, another common empirical band structure methods for semiconductors.

\section{Methods and Theory}
\label{II}
In this section standard k.p Hamiltonians that represent bulk zincblende and rhombohedral crystal structures are briefly discussed. These bulk Hamiltonians describe the dispersion relationship around the high-symmetry $ \Gamma $ point. Bi$_{2}$Te$_{3}$, Bi$_{2}$Se$_{3}$, and Sb$_{2}$Te$_{3}$ which exhibit topological insulator behaviour are well-known examples of compounds with a rhombohedral structure while common semiconductors such as GaAs and InAs belong to the zincblende family.
\subsection{Representative Hamiltonians for zincblende and rhombohedral crystal}
The theory of Luttinger and Kohn uses an angular momentum-dependent basis functions for constructing the eight-band k.p Hamiltonian for zincblende crystal. The basis states for constructing this Hamiltonian are given in appendix A. The Hamiltonian~\cite{liu2002modeling,chuang2012physics} $ \left( H_{8 \times 8} \right) $ is constructed by including the heavy hole, light hole, spin orbit split off valence band, and the lowest conduction band. The valence and conduction bands are linearly coupled. Further, it is assumed that the \textit{z}-axis of the crystal is aligned along $\left[001\right] $ axis. The k.p Hamiltonian can also be transformed~\cite{park2000comparison} for an arbitrary crystal growth axis.

The four-band Hamiltonian~\cite{zhang2009topological} for Bi$_{2}$Te$_{3}$, Bi$_{2}$Se$_{3}$, and Sb$_{2}$Te$_{3}$ is constructed (Eq.~\ref{eqn1}) in terms of the four lowest low-lying states $ \vert P1_{z}^{+} \uparrow \rangle $, $ \vert P2_{z}^{-} \uparrow \rangle $, $ \vert P1_{z}^{+} \downarrow \rangle $, and $ \vert P2_{z}^{-} \downarrow \rangle $. These materials have assumed significance as topological insulators. Topological insulators~\cite{hasan2010colloquium,qi2011topological,ando2013topological} are a new class of materials whose surfaces host bound Dirac fermion like spin-polarized particles with high mobility. These states, in a time reversal invariant system~\cite{kane2005quantum,roy2009topological} are protected against perturbation and non-magnetic disorder. 

\begin{align}
\label{eqn1}
H(k) = \epsilon(k) + \begin{pmatrix}
M(k) & A_{1}k_{z} & 0 & A_{2}k_{-} \\
A_{1}k_{z} & -M(k) & A_{2}k_{-} & 0 \\
0 & A_{2}k_{+} & M(k) & -A_{1}k_{z} \\
A_{2}k_{+} & 0 & -A_{1}k_{z} & -M(k) \\
\end{pmatrix}
\end{align}
where $ \epsilon(k) = C + D_{1}k_{z}^{2} + D_{2}k_{\perp}^{2}$, $ M(k) = M_{0} + B_{1}k_{z}^{2} + B_{2}k_{\perp}^{2}$ and $ k_{\pm} = k_{x} \pm ik_{y}$. For Bi$_{2}$Te$_{3}$ and Bi$_{2}$Se$_{3}$, the relevant parameters are summarized in Table.~\ref{table1}.
\begin{widetext}
\begin{equation}
H_{8 \times 8} =  \begin{pmatrix} 
E_{c}+A & -\sqrt{3}\nu & \sqrt{2}U & U & 0 & 0 & \nu^{*} & \sqrt{2}\nu^{*} \\
-\sqrt{3}\nu^{*} & E_{v}-P-Q & S & \dfrac{1}{\sqrt{2}}S & 0 & 0 & -R & -\sqrt{2}R \\
\sqrt{2}U & S^{*} & E_{v}-P+Q & \sqrt{2}Q & -\nu^{*} & -R & 0 & -\sqrt{\dfrac{3}{2}}S \\
U & \dfrac{1}{\sqrt{2}}S^{*} & \sqrt{2}Q &  E_{v}-P-\Delta  & \sqrt{2}\nu^{*} & \sqrt{2}R & -\sqrt{\dfrac{3}{2}}S & 0 \\
0 & 0 & -\nu & \sqrt{2}\nu & E_{c}+A & \sqrt{3}\nu^{*} &  \sqrt{2}U & -U \\
0 & 0 & -R^{*} & \sqrt{2}R^{*} & \sqrt{3}\nu &  E_{v}-P-Q & -S^{*} & \dfrac{1}{\sqrt{2}}S \\ 
\nu & -R^{*} & 0 & -\sqrt{\dfrac{3}{2}}S^{*} & \sqrt{2}U & -S & E_{v}-P+Q  & -\sqrt{2}Q \\
\sqrt{2}\nu & -\sqrt{2}R^{*} & -\sqrt{\dfrac{3}{2}}S^{*} & 0 & -U & \dfrac{1}{\sqrt{2}}S & -\sqrt{2}Q &  E_{v}-P-\Delta 
\label{zbham}
\end{pmatrix}
\end{equation}
\end{widetext} 
where 
\begin{gather}
A = \left(\dfrac{\hbar^{2}}{2m^{*}_{e}}\right) \left( k_{t}^{2}+k_{z}^{2}\right) \notag \\
P = \left(\dfrac{\hbar^{2}}{2m^{*}_{0}}\right)\gamma_{1}\left( k_{t}^{2}+k_{z}^{2}\right) \notag \\
Q = \left(\dfrac{\hbar^{2}}{2m^{*}_{0}}\right)\gamma_{2}\left( k_{t}^{2}-2k_{z}^{2}\right) \notag \\ 
R = \left(\dfrac{\hbar^{2}}{2m^{*}_{0}}\right)\sqrt{3}\left[-\gamma_{2}\left( k_{x}^{2}-k_{y}^{2}\right)+2i\gamma_{3}k_{x}k_{y}\right] \notag \\
S = \left(\dfrac{\hbar^{2}}{2m^{*}_{0}}2\right)\sqrt{3}\gamma_{3}\left(k_{x}-ik_{y}\right)k_{z} \notag \\
\nu = \dfrac{1}{\sqrt{6}}P_{cv}\left(k_{x} + ik_{y}\right) \notag \\
U = \dfrac{1}{\sqrt{3}}P_{cv}k_{z} \notag \\
P_{cv} = \dfrac{\hbar}{m_{0}}\langle iS\vert \dfrac{\hbar}{i}\dfrac{\partial}{\partial x}\vert X\rangle = \sqrt{\left(\dfrac{\hbar^{2}}{2m^{*}_{0}}\right)E_{p}}
\label{symb}
\end{gather} 
$ \gamma_{1}, \gamma_{2},$ and $ \gamma_{3}$ are Kohn-Luttinger parameters. $ m_{e}^{*} $ is the effective mass of conduction band electron at $ \Gamma $ point and $ m_{0}^{*}$ is the free electron mass. $ E_{c}$ and $ E_{v}$ are the unstrained conduction and valence band edges, respectively. $ \Delta $ defines the spin-orbit splitting. All parameters for III-V materials used in this work are obtained from the paper by Vurgaftman, Meyer, and Ram-Mohan.~\cite{vurgaftman2001band}
\begin{table}[htb]
\caption{4-band k.p parameters~\cite{liu2010model} for Bi$_{2}$Te$_{3}$ and Bi$_{2}$Se$_{3}$.}
\centering
\label{table1}
\begin{tabular}{lcc}
\noalign{\smallskip} \hline \hline \noalign{\smallskip}
Parameters & Bi$_{2}$Te$_{3}$ & Bi$_{2}$Se$_{3}$ \\\hline
M$_{0}$ (eV) & 0.30 & 0.28 \\
A$_{1}$ (eV \AA ) & 2.26 & 2.2 \\
A$_{2}$ (eV \AA )  & 2.87 & 4.1 \\
B$_{1}$ (eV \AA$^{2}$)  & 10 & 10 \\
B$_{2}$ (eV \AA$^{2}$)  & 57.38 & 56.6 \\
C (eV) & -0.18 & -0.0068 \\
D$_{1}$ (eV \AA$^{2}$) & 6.55 & 1.3 \\
D$_{2}$ (eV \AA$^{2}$) & 29.68 & 19.6 \\
\noalign{\smallskip} \hline \noalign{\smallskip}
\end{tabular}
\end{table} 

\section{Finite difference discretization of a k.p Hamiltonian}
\label{III}
Numerical diagonalization of the k.p Hamiltonian can be achieved by transforming the linear Schr{\"o}dinger equation $ H\Psi = E_{n}\Psi $ into an equivalent matrix representation. As a first step, the domain of interest is partitioned into a series of interconnected nodes generated by utilizing methods available in the finite element~\cite{ramdas2002finite} and finite difference procedures. Subsequently, an aggregation of the Schr{\"o}dinger equation defined at each node sets up the matrix representation. In this section, the most general finite difference discretized version of a Hamiltonian is presented which can be further extended to any k.p Hamiltonian of arbitrary size. To begin, a Hamiltonian in its most complete discretized form can be written as 
\begin{align}
\begin{split}
H_{0} = \sum\limits_{i=x,y,z}A_{i}k_{i}^{2} +  \sum\limits_{i=x,y,z}B_{i}k_{i} +  \dfrac{1}{2}\sum\limits_{i,j = x,y,z}C_{i}k_{i}k_{j} \\
+ Const
\label{coeff_matrix}
\end{split}
\end{align}
The $ Const $ includes energy band-gap, band offsets, and spin-orbit induced splitting. For the eight-band zincblende Hamiltonian used in this work, it takes a simple form of a matrix of size $ 8 \times 8 $ and populates only the leading diagonal terms shown in Eq.(~\ref{diag}).
\begin{equation}
Const = \left\lbrace  E_{c}, E_{v}, E_{v}, E_{v}-\Delta, E_{c}, E_{v}, E_{v},E_{v}-\Delta\right\rbrace 
\label{diag}
\end{equation}

Each of the matrices belonging to the $ A_{i}, B_{i}$, and $ C_{i} $ group is of size $ n \times n $ where $ n $ denotes the number of basis functions. For instance, the bulk zincblende Hamiltonian is built of eight angular-momentum dependent orbitals; $ n $ is therefore equal to eight in this case. Assuming no periodicity, such as a quantum dot, the $ \overrightarrow{k} $ vector can be replaced by the standard differential operator notation $ \overrightarrow{k} = -\sqrt{-1}\dfrac{\partial}{\partial i} $ where $ i = \left\lbrace x,y,z\right\rbrace $. This turns the Schr{\"o}dinger equation (Eq.(~\ref{coeff_matrix})) into a series of differential operators such as $ A_{i}\dfrac{\partial^{2}}{\partial i^{2}} $ and  $ B_{i}\dfrac{\partial}{\partial i} $.  

The differential operators $ A_{i}\dfrac{\partial^{2}}{\partial i^{2}} $ and  $ B_{i}\dfrac{\partial}{\partial i} $ are defined as follows in the finite difference approximation.~\cite{chuang1997band} The actual analytic forms of $ A_{i} $ and $ B_{i} $, and $ C_{i} $ depend on the chosen Hamiltonian representation. Representative matrices of size $ n \times n $ for a zincblende eight-band Hamiltonian is derived in Appendix B. Steps to discretize the mixed differential operators such as $ C_{i}\dfrac{\partial^{2}}{\partial x_{i} \partial x_{j}} $ are worked out later.
\begin{subequations}
\begin{gather}
A(z)\dfrac{\partial^{2}\psi}{\partial z^{2}} \longrightarrow \dfrac{\partial}{\partial z}\left(A(z)\dfrac{\partial \psi}{\partial z}\right)\bigg|_{z = z_{i}} \notag \\
\approx \dfrac{A\left( z_{i+1}\right)+ A\left( z_{i}\right)}{2 \left(\Delta z\right)^{2}}  \psi \left(z_{i+1}\right) \notag \\
-\dfrac{A\left( z_{i-1}\right)+ 2A\left( z_{i}\right)+ A\left( z_{i+1}\right) }{2 \left(\Delta z\right)^{2}}  \psi \left(z_{i}\right) \notag \\
+\dfrac{A\left( z_{i}\right)+ A\left( z_{i-1}\right)}{2 \left(\Delta z\right)^{2}}  \psi \left(z_{i-1}\right) \notag \label{ddsc}\\
\end{gather}
and
\begin{gather}
B(z)\dfrac{\partial\psi}{\partial z} \longrightarrow \dfrac{1}{2}\left(B(z)\dfrac{\partial \psi}{\partial z}  + \dfrac{\partial \left( B\psi\right) }{\partial z}\right)\bigg|_{z = z_{i}} \notag \\
\approx \dfrac{B\left( z_{i+1}\right)+ B\left( z_{i}\right)}{4 \left(\Delta z\right)}  \psi \left(z_{i+1}\right)- \dfrac{B\left( z_{i}\right)+ B\left( z_{i-1}\right)}{4 \left(\Delta z\right)}  \psi \left(z_{i-1}\right)\notag \label{sdsc}\\
\end{gather}
\end{subequations}
where $\Delta z $ is size of the mesh along \textit{}z-axis in the finite difference grid. The mesh size for discretization along $ {x,y,z} $ axes can be set to arbitrary values, subject to numerical accuracy.

The finite difference discretization of differential operators which include such cross terms using the method shown in Eq.(~\ref{ddsc}, ~\ref{sdsc}) is given below. For purpose of illustration, let \textit{P} denote any arbitrary matrix such that the differential operator $ \dfrac{\partial^{2} P}{\partial x_{i}x_{j}} $ needs to be discretized
\begin{align}
\dfrac{\partial^{2}}{\partial x_{i}x_{j}}\left(P\Psi\right) = \dfrac{1}{2}\left[ \dfrac{\partial}{\partial x_{i}}P \dfrac{\partial \Psi}{\partial x_{j}} 
+ \dfrac{\partial}{\partial x_{j}}P \dfrac{\partial \Psi}{\partial x_{i}}\right] 
\label{cross_term}
\end{align} 
Let $ f $ denote $ P \dfrac{\partial \Psi}{\partial x_{j}} $. The first half of Eq(~\ref{cross_term}) therefore can be written as $ \dfrac{\partial}{\partial x_{i}} P \dfrac{\partial \Psi}{\partial x_{j}} = \dfrac{\partial f}{\partial x_{i}}  $. This expression can be further expanded as shown below
\begin{subequations}
\begin{align}
\begin{split}
\dfrac{\partial f}{\partial x_{i}} = \dfrac{1}{2}\left[\dfrac{f_{i+1} - f_{i-1}}{2a}\right] 
\end{split}
\label{crfs}
\end{align}
Using the definition of $ f $ as expressed above, each term of Eq(~\ref{crfs}) further expands to 
\begin{align}
\begin{split}
f_{i+1} = P_{i+1,j}\left[\dfrac{\Psi_{i+1,j+1} - \Psi_{i+1,j-1}}{2a}\right] \\
f_{i-1} = P_{i-1,j}\left[\dfrac{\Psi_{i-1,j+1} - \Psi_{i-1,j-1}}{2a}\right]
\end{split}
\label{crss}
\end{align}
Combining all the terms and writing the full expression one obtains
\begin{align}
\begin{split}
\dfrac{1}{2}\left[ \dfrac{\partial}{\partial x_{i}}P \dfrac{\partial \Psi}{\partial x_{j}}\right]  = \dfrac{1}{8a^{2}}\left[P_{i+1,j}\left\lbrace \Psi_{i+1,j+1} - \Psi_{i+1,j-1}\right\rbrace \right]  \\
- \left[ P_{i-1,j}\left\lbrace \Psi_{i-1,j+1} - \Psi_{i-1,j-1}\right\rbrace \right] 
\end{split}
\label{crts}
\end{align}
Similarly, the other half of Eq(~\ref{cross_term}) can be written following an identical procedure as
\begin{align}
\begin{split}
\dfrac{1}{2}\left[ \dfrac{\partial}{\partial x_{j}}P \dfrac{\partial \Psi}{\partial x_{i}}\right]  = \dfrac{1}{8a^{2}}\left[P_{i,j+1}\left\lbrace \Psi_{i+1,j+1} - \Psi_{i-1,j+1}\right\rbrace \right]  \\
- \left[ P_{i,j-1}\left\lbrace \Psi_{i+1,j-1} - \Psi_{i-1,j-1}\right\rbrace \right] 
\end{split}
\label{crffs}
\end{align}
\end{subequations} 

Each node $ {x_{i}, y_{i}, z_{i}} $, in the current finite difference scheme, gives rise to a set of eighteen neighbours. The neighbours for a given node \textit{x,y,z} can be enumerated as $ (x\pm a,y,z), (x,y\pm a,z), (x,y,z\pm a), (x\pm a,y\pm a,z), (x, y\pm a, z\pm a), (x\pm a, y, z\pm a) $. Each of these neighbours contributes a coupling matrix to the node under consideration. Neighbours that do not belong to any discretized on the finite difference grid are assumed to have a zero contribution. Such neighbours typically arise for nodes along the edge of the finite-difference grid. Additionally, the wave function $ \psi $ is set to zero at the grid boundaries. The grid is thus a 3D-cubic representation of any device. In interest of brevity, a neighbour, for instance, $ (x\pm a,y,z)$ is succinctly expressed as $ (i \pm 1, y, z) $ later in the text where $ (i,j,k) $ represent as usual the chosen node in the domain.

\subsection{Analytic expressions for coupling matrices}
Each of the eighteen neighbours described above couple to the chosen node $ {i,j,k} $ through a matrix ($ M_{p,q,r}$ ) of size $ n \times n $ where $ n $ denotes the number of basis functions and $ p,q $, and $ r $ represent the neighbour coordinates. The algebraic expression for each coupling  matrix is worked out below. To translate the expressions given into code form, one needs to set up a matrix of size  $ N \times N $ where $ N = n \times N_{x}\times N_{y}\times N_{z} $. $ N_{x},N_{y}$, and $ N_{z} $ designate the number of discretized points along the \textit{x,y}, and \textit{z} axes respectively. For a given node located in this matrix with a specific row and column index, the coupling matrices are arranged by placing them either on left or right of the node. It is important to note that the row index of all the coupling matrices are identical to that of the chosen node, the column index is suitably varied to place them correctly on $ H_{N \times N}$ which represents the full device Hamiltonian. 

Let the coefficient of $ k_{x}^{2}, k_{y}^{2} $, and $ k_{z}^{2} $ be denoted as $ A_{1}, A_{2} $, and $ A_{3} $ respectively. The matrix that describes the self-coupling or the diagonal block of size $ n \times n $ is noted in Eq.(~\ref{self_mat}) below. Throughout the derivation of coupling matrices, it is tacitly assumed that the finite-difference discretization along each of the three axes is a constant and set equal to \textit{a} in arbitrary units.
\begin{align}
\begin{split}
M_{i,j,k}^{self} = \dfrac{A_{1}(i-1,j,k) + 2A_{1}(i,j,k) + A_{1}(i+1,j,k)}{2a^{2}} \\
+ \dfrac{A_{2}(i,j-1,k) + 2A_{2}(i,j,k) + A_{2}(i,j+1,k)}{2a^{2}}  \\
+ \dfrac{A_{3}(i,j,k-1) + 2A_{3}(i,j,k) + A_{1}(i,j,k+1)}{2a^{2}} 
\end{split}
\label{self_mat} 
\end{align}
The construction of Eq.(~\ref{self_mat}) can be easily verified by collecting the coefficients of wave function $ \Psi_{i,j,k} $ in Eq.(~\ref{ddsc}). $ \Psi_{i,j,k} $ is the wave function component that \textit{belongs} to node $ (i,j,k) $. Note that the complete set of wave functions for the device Hamiltonian is a column vector of size $ N \times 1 $. Further, wave function at each node is constituted of $ n $ sub-components. 

The coupling block for the eighteen neighbours are established below following the same discretization scheme as above. Let the coefficient of $ k_{x}, k_{y} $, and $ k_{z} $ be denoted as $ B_{1}, B_{2} $, and $ B_{3} $ respectively. The two neighbours situated on the discretized \textit{x}-axis can be written as shown in Eq.(~\ref{xnbr}). Note that \textit{imag} stands for the imaginary quantity $ \sqrt{-1} $ and should not be confused with $ i $ in the triad $ \left\lbrace i,j,k \right\rbrace $ which simply represents a set of coordinates in discretized space.
\begin{align}
\begin{split}
M_{i\pm 1,j,k} = -\dfrac{A_{1}(i\pm 1,j,k) + A_{1}(i,j,k)}{2a^{2}} \\
\mp imag\dfrac{B_{1}(i,j,k) +  B_{1}(i\pm 1,j,k)}{4a} 
\end{split}
\label{xnbr} 
\end{align}
As before, this coupling matrix is developed by collecting all coefficients of wave function $ \Psi_{i \pm 1,j,k} $ obtained from the discretization steps Eq(~\ref{ddsc}). Similar expressions can be constructed for neighbours situated on the discretized \textit{y} and \textit{z}-axis 
\begin{subequations}
\begin{align}
\begin{split}
M_{i,j \pm 1,k} = -\dfrac{A_{2}(i,j\pm 1,k) + A_{2}(i,j,k)}{2a^{2}}  \\
\mp imag\dfrac{B_{2}(i,j,k) +  B_{2}(i,j\pm 1,k)}{4a}  \\
\end{split}
\label{ynbr}
\end{align}
\begin{align}
\begin{split}
M_{i,j ,k\pm 1} = -\dfrac{A_{3}(i,j,k\pm 1) + A_{3}(i,j,k)}{2a^{2}}  \\
\mp imag\dfrac{B_{3}(i,j,k) +  B_{3}(i,j,k\pm 1)}{4a} 
\end{split}
\label{znbr}
\end{align}
\end{subequations}
The last set of neighbours are now introduced which involve matrices which are coefficients of the cross terms $ k_{i}k_{j} $ where $ i,j = x,y,z $. Using expressions derived in Eq(~\ref{crts})and Eq(~\ref{crffs}), they can now be written as follows. This group of matrices is identified with a prime.
\begin{subequations}
\begin{align}
M_{i+1,j \pm 1,k}^{'} = \mp\left[\dfrac{C_{1}(i+1,j,k) + C_{1}(i,j \pm 1,k)}{8a^{2}}\right] 
\label{ne}
\end{align}
For coupling matrices $ M_{i-1,j \pm 1,k}^{'} $, the expressions evaluate to
\begin{align}
M_{i-1,j \pm 1,k}^{'} = \pm\left[\dfrac{C_{1}(i-1,j,k) + C_{1}(i,j \pm 1,k)}{8a^{2}}\right] 
\label{sw}
\end{align}
Matrices $ M_{i,j+1,k \pm 1}^{'} $ that represent neighbours $ (i,j+1,k \pm 1) $ are given as
\begin{align}
M_{i,j+1,k \pm 1}^{'} = \mp\left[\dfrac{C_{2}(i,j+1,k) + C_{2}(i,j,k \pm 1)}{8a^{2}}\right] 
\label{hwn}
\end{align}
Likewise, matrices $ M_{i,j-1,k \pm 1}^{'} $ representing neighbours $ (i,j-1,k \pm 1) $ can be written as
\begin{align}
M_{i,j-1,k \pm 1}^{'} = \pm\left[\dfrac{C_{2}(i,j-1,k) + C_{2}(i,j,k \pm 1)}{8a^{2}}\right] 
\label{hws}
\end{align}
The final set of coupling blocks for neighbours $ M_{i+1,j,k \pm 1}^{'} $ and $ M_{i-1,j,k \pm 1}^{'} $ are
\begin{align}
M_{i+1,j,k \pm 1}^{'} = \mp\left[\dfrac{C_{3}(i+1,j,k) + C_{2}(i,j,k \pm 1)}{8a^{2}}\right] 
\label{hwe}
\end{align}
and
\begin{align}
M_{i-1,j,k \pm 1}^{'} = \pm\left[\dfrac{C_{3}(i-1,j,k) + C_{2}(i,j,k \pm 1)}{8a^{2}}\right] 
\label{hww}
\end{align}
\end{subequations} 
All the coupling blocks have now been identified to set up the full Hamiltonian. If the Hamiltonian represents a heterostructure, the coupling matrices are material dependent. It is therefore necessary to allow the code to determine the respective material domain in which the node $ (i,j,k) $ and its eighteen neighbours lie. The number of neighbours will reduce when nodes on boundary of the discretized device are considered.

The equivalent discretized representation for a nanowire is simpler as shown below.
\begin{align}
\begin{split}
H_{wire} = \sum\limits_{i=x,y}A_{i}k_{i}^{2} +  \sum\limits_{i=x,y}B_{i}k_{i} +  \dfrac{1}{2}\sum\limits_{i,j = x,y}C_{i}k_{i}k_{j} \\
+ Const
\label{coeff_matrixwire}
\end{split}
\end{align}
This nanowire is assumed to be dimensionally confined along the \textit{x,y}-axes and periodic along the \textit{z}-axis. The $\overrightarrow{k_{x}} $ and $\overrightarrow{k_{y}} $ vectors are not therefore not good quantum numbers and converted in to operator notation(Eq.(~\ref{coeff_matrixwire})). Coefficient matrices of the $\overrightarrow{k_{x}} $ are treated as a part of the $ Const $ and do not take part in the discretization process. The $ Const $ matrix for a wire with the stated confinement takes a very different structure compared to its equivalent representation (Eq.(~\ref{diag}) in a fully discretized form. The $ Const $ matrix for a wire is given in Appendix B.

\section{Continuum strain model}
\label{IV}
The algorithm~\cite{pryor1998eight,pryor1999geometry} for setting up a matrix equation to perform strain calculations is described here. It is assumed that there are \textit{N$_{x}$}, \textit{N$_{y}$}, and \textit{N$_{z}$} points or nodes along the \textit{x,y,z} axes respectively. The displacement vector for each node is initially set to zero. An intrinsic strain vector(6 $\times$ 1) is created and all points in the substrate have zero entries. Every other node on the discretized device will have an intrinsic strain defined by Eq.~\ref{int_strain} 

\begin{equation}
\varepsilon_{xx} = \dfrac{a_{lattice} - a_{sub}}{a_{sub}} = \varepsilon_{yy}
\label{int_strain}
\end{equation}
where \textit{a$_{sub}$} and \textit{a$_{lattice}$} are the lattice constants of the substrate and any other material constituent of the device in which the current node of interest lies respectively. For a zincblende crystal, $ \varepsilon_{zz} $ can be simply computed using the Poisson ratio and $ \varepsilon_{xx} $.

A 6 $\times$ 3 matrix is now constructed that will operate on the 3 $\times$ 1 displacement vector to produce a 6 $\times$ 1 strain vector. The 6 $\times$ 3 matrix, \textit{D} is defined as follows

\begin{equation}
D = \begin{pmatrix}
\dfrac{\partial}{\partial x} && 0 && 0 \\
0 && \dfrac{\partial}{\partial y} && 0 \\
0 && 0 && \dfrac{\partial}{\partial z} \\
\dfrac{1}{2}\dfrac{\partial}{\partial y} && \dfrac{1}{2}\dfrac{\partial}{\partial x} && 0 \\
 && && \\
0 && \dfrac{1}{2}\dfrac{\partial}{\partial z} && \dfrac{1}{2}\dfrac{\partial}{\partial y} \\
 && && \\
\dfrac{1}{2}\dfrac{\partial}{\partial z} && 0 && \dfrac{1}{2}\dfrac{\partial}{\partial x}
\end{pmatrix}
\end{equation}      
The matrix product of \textit{D} and a 3 $\times$ 1 displacement vector \textit{u} produces the 6 $\times$ 1 strain vector in Voigt notation. The elastic energy of a zincblende or cubic crystal is given as:
\begin{equation}
\begin{split}
E = \dfrac{1}{2}C_{11}(\varepsilon_{xx}^{2}+\varepsilon_{yy}^{2}+\varepsilon_{zz}^{2})+C_{12}(\varepsilon_{xx}\varepsilon_{yy}+ \varepsilon_{yy}\varepsilon_{zz}+\varepsilon_{zz}\varepsilon_{xx}) \\
+ 2C_{44}(\varepsilon_{xy}^{2}+\varepsilon_{yz}^{2}+\varepsilon_{zx}^{2})  
\label{elasticE}
\end{split}
\end{equation} 
The elastic energy equation (Eq.~\ref{elasticE}) can be expressed in matrix form as:
\begin{equation}
E = u^{T}D^{T}GDu
\label{elasticE_matrix}
\end{equation}
where \textit{G} is the representative elastic constant matrix for a given crystal system. In particular, for cubic crystals it takes the form below (Eq.~\ref{elasticmatrix}). In calculations below, the size of \textit{G} matrix is $ 6N_{p} \times 6N_{p}$ where $ N_{p} = N_{x} \times N_{y} \times N_{z} $ and the \textit{D} matrix is $ 6N_{p} \times 3N_{p}$  

\begin{equation}
G = \begin{pmatrix}
C_{11} & C_{12} & C_{12} & 0 & 0 & 0 \\
C_{12} & C_{11} & C_{12} & 0 & 0 & 0 \\
C_{12} & C_{12} & C_{11} & 0 & 0 & 0 \\
0 & 0 & 0 & C_{44} & 0 & 0 \\
0 & 0 & 0 & 0 & C_{44} & 0 \\
0 & 0 & 0 & 0 & 0 & C_{44} 
\end{pmatrix}
\label{elasticmatrix}
\end{equation}
The final step is then to evaluate $ \dfrac{\partial E}{\partial u_{i}} $ where \textit{i = x,y,z}
\begin{equation}
\dfrac{\partial E}{\partial u_{i}} = \dfrac{\partial}{\partial u_{i}}\left[\left( u^{T}D^{T}+\varepsilon_{0}\right)G\left( uD+\varepsilon_{0}\right) \right]
\label{diff1}
\end{equation}
where $\varepsilon_{0}$ is the intrinsic strain.
Expanding (~\ref{diff1}), we have
\begin{equation}
\begin{split}
\dfrac{\partial E}{\partial u_{i}} = \dfrac{\partial}{\partial u_{i}}\left[ u^{T}D^{T}GDu + u^{T}D^{T}G\varepsilon_{0}
+\varepsilon_{0}^{T}GDu+\varepsilon_{0}^{T}G\varepsilon_{0} \right]
\end{split}
\label{diff2} 
\end{equation} 
Dropping the subscript \textit{i} and defining $D^{T}GD$ as \textit{A}, Eq(~\ref{diff2}) can be written as
\begin{equation}
\begin{split}
\dfrac{\partial E}{\partial u_{i}} = \dfrac{\partial u^{T}}{\partial u}Au + u^{T}\dfrac{\partial A}{\partial u}u + u^{T}A\dfrac{\partial u}{\partial u} + \\
\dfrac{\partial u^{T}}{\partial u}D^{T}G\varepsilon_{0} + u^{T}\dfrac{\partial D^{T}}{\partial u}G\varepsilon_{0} \\
+ \varepsilon_{0}^{T}GD\dfrac{\partial u}{\partial u} + \dfrac{\partial\left( \varepsilon_{0}^{T}G\varepsilon_{0}\right)}{\partial u} \\
\label{diff3}
\end{split}
\end{equation}

Simplifying and noting that the second, fifth, and seventh terms evaluate to zero, Eq.(~\ref{diff3})is now reduced to
\begin{equation}
\left[0....1...0\right]Au + u^{T}A\left[0....1...0\right]^{T} = -\left(D^{T}G\varepsilon_{0} + \varepsilon_{0}^{T}GD\right) 
\label{diff4}
\end{equation}
To obtain displacement of each node, the energy derivative $ \dfrac{\partial E}{\partial u_{i}} $ is set to zero. Collecting terms, the final equation is
\begin{equation}
Au = - D^{T}G\varepsilon_{0}
\label{diff5}
\end{equation}
Note that Eq(~\ref{diff5}) is a linear equation where each term is now reduced to a number from the original matrix form. Carrying out the same operation on each node, a set of linear equations is obtained which must now be solved to get the final displacement and the associated 6 $\times$ 1 strain tensor. The strain vector is used to construct the Bir-Pikus strain Hamiltonian~\cite{henderson1996effective} and couple to the k.p Hamiltonian described above.

\section{Results}
\label{V}
This section brings together all the methods collected above. Two cases are considered 1) A quantum dot heterostructure and 2) A topological insulator slab.
\subsection{Quantum Dot Heterostructure}
\begin{figure}
\includegraphics[scale= 0.8]{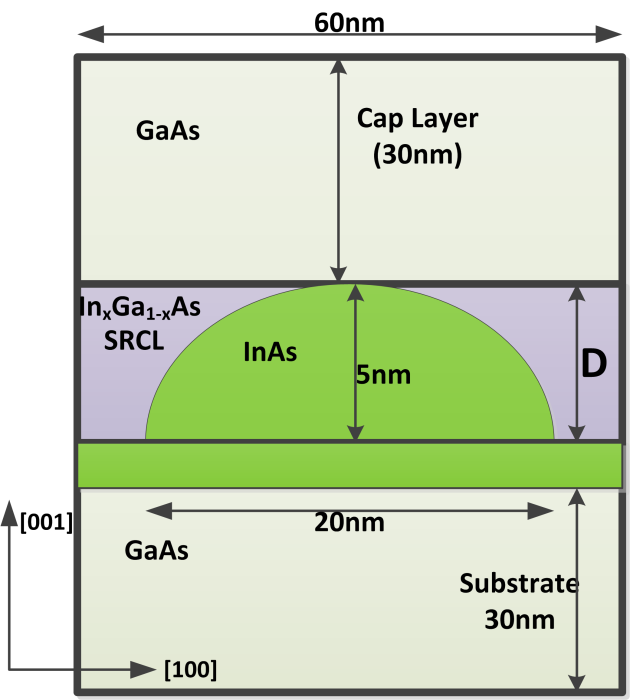}
\caption{Cross section through the modeled self-assembled quantum dot. The InAs dot itself is dome-shaped and embedded in an InGaAs stress-reducing layer (SRCL) with mole fraction x=0.4. The structure is embedded in a GaAs matrix.} 
\label{fig1}
\end{figure}

\begin{figure}
\includegraphics[scale= 0.7]{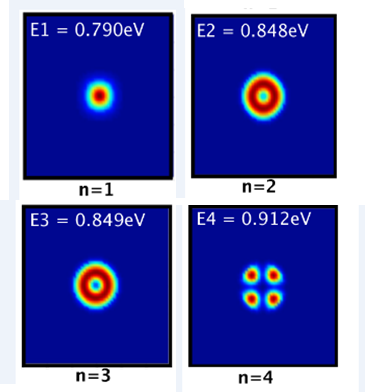}
\caption{Wave functions and eigen states for $ n = 1-4 $ from an eight-band k.p calculation. The radially symmetric wave functions for $ n = 2, 3 $ are an outcome of the assumed C$_{4v}$ point group symmetry of the zincblende k.p Hamiltonian instead of the actual C$_{2v}$.}
\label{fig2}
\end{figure}

Figure ~\ref{fig1} shows the schematic of the simulated system. A 10 {\AA} dome-shaped InAs QD of 5 $\mathrm{nm}$ height and 20 $\mathrm{nm}$ base diameter sits on a single monolayer of InAs that serves as the wetting layer. The QD is set in an InxGa1-xAs alloy of mole fraction 40$\%$ which functions as the stress-reducing layer (SRCL). The height of the SRCL is 5.0 $ \mathrm{nm}$. A GaAs host matrix (60 $\mathrm{nm}$ x 60 $ \mathrm {nm}$ x 60 $ \mathrm{nm}$) surrounds the whole structure. 
The first four wave functions and eigen states of this dot are shown in Fig.~\ref{fig2}. The lowest eigen state is at 0.790 $ \mathrm{eV} $ for $ n = 1 $ and is equal to 0.912 $ \mathrm{eV} $ for $ n = 4 $. The shape of the wave function for the ground state $ n = 1 $ corresponds to an \textit{s}-like state and is therefore spherical in nature. For $ n = 2, 3 $, the \textit{p}-type orbitals come to play and in keeping with the overall C$_{4v} $ group symmetry of the k.p Hamiltonian~\cite{mildred2008group}, the wave functions exhibit a symmetric shape though in principle they are split.~\cite{sengupta2011multiscale} Strain has not been included in these calculations. Strain-induced deviations are presented later.

\begin{figure}
\includegraphics[scale= 0.4]{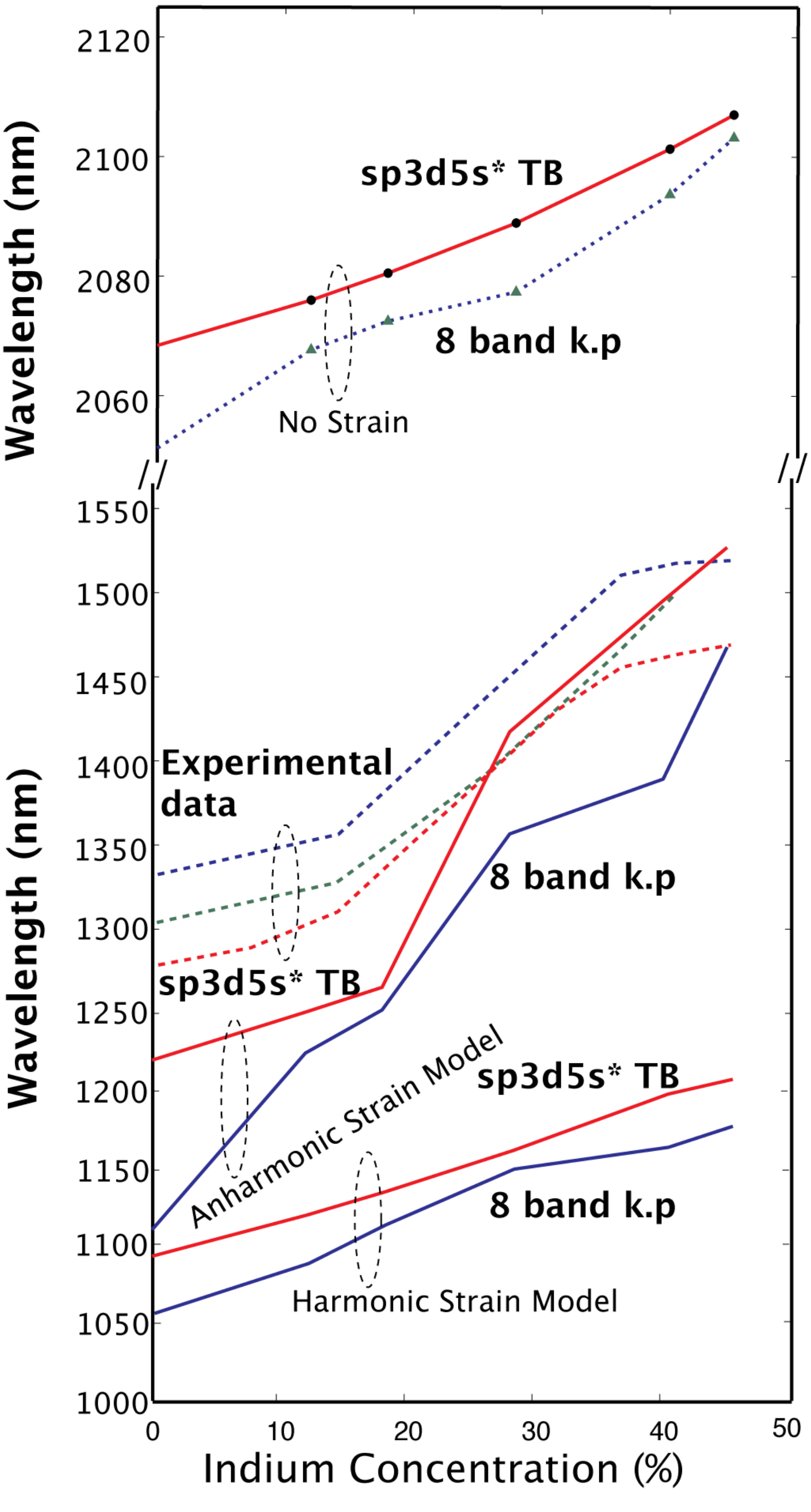}
\caption{Optical Transition wavelength vs. Indium conc. A much closer match is obtained with experimental data for an anharmonic VFF model combined with the 8-band continuum k.p.} 
\label{fig3}
\end{figure}

While unstrained calculations validate theoretical expectations arising out of the known symmetry of the k.p Hamiltonian, intrinsic strain is inherent in such quantum dot heterostructures and introduce significant changes to the eigen states. Changed eigen states eventually affect the optical transition rates. Strain is therefore incorporated in these calculations in two ways: 1)Using the continuum strain model introduced in Section~\ref{IV} and 2)Interpolation of anharmonic VFF to the k.p Hamiltonian.  Interpolation of VFF strain from an atomistic basis to continuum grid is performed by constructing a sphere around each continuum node. An average value for each of the six strain component is computed by considering all atoms from the atomistic grid that lie within this sphere. The average strain tensor components are then used in a standard Bir-Pikus Hamiltonian. As a comparison, data obtained through tight binding calculations~\cite{usman2009moving}for both harmonic and anharmonic VFF models have been shown in Table.~\ref{table2} and Fig.~\ref{fig3}. It is seen that a while a simulation under a strain model with a harmonic approximation gets the results closer to the laboratory data, the neglect of anharmonic modifications of the inter-atomic potential leaves ample room for improvement. Anharmonicity is added to the inter-atomic potential by utilizing distance- and angle-dependent VFF constants~\cite{lazarenkova2004effect}. By inclusion of anharmonic corrections in the simulation, the results are within 4-9 $\%$ of reported experimental values~\cite{tatebayashi2001over} (see Table~\ref{table2} and Fig.~\ref{fig4}). Another fact supporting the inclusion of anharmonicity, a non-linear dependence of the emission wavelengths as a function of the SRCL mole fraction which is in agreement with experiment is not discussed here. Non-linearity inherent in the atomistic description is not captured well by continuum k.p as borne out by Fig.~\ref{fig4}. Further corrections due to piezoelectric effects operational in polar GaAs were not considered in this work.
\begin{table}
\caption{Transition energies between the first confined electron and hole states for various models and experimental values.}
\centering
\label{table2}
\begin{tabular}{lcc}
\noalign{\smallskip} \hline \hline \noalign{\smallskip}
Model & E-H transition energy ($\mathrm{eV}$) \\\hline
Experiment & 0.81 - 0.85\\
k.p unstrained  & 0.594 \\
TB unstrained & 0.591 \\
k.p strained (harmonic continuum)  & 1.079 \\
TB strained (harmonic VFF)  & 1.040 \\
k.p strained (anharmonic VFF) & 0.885 \\
TB strained (anharmonic VFF) & 0.828 \\
\noalign{\smallskip} \hline \noalign{\smallskip}
\end{tabular}
\end{table}

\subsection{Topological Insulators}
The four-band k.p Hamiltonian for topological insulators is discretized in the usual fashion described in Section.~\ref{III}. It is worthwhile to note that since the dispersion of a quantum well (confinement along \textit{z}-axis is desired, the two differential operators needed are $ -P_{z^{2}}\dfrac{\partial^{2}}{\partial z_{2}} $ and $ -iQ_{z}\dfrac{\partial}{\partial z} $, where $ P_{z^{2}} $ and $ Q_{z} $ are the coefficients of  $ k^{2}_{z} $ and $ k_{z}  $ respectively. All other terms are represented through a matrix that is a function of $ k_{x} $ and $ k_{y}$ in addition to other material-dependent constants present in the Hamiltonian.

\begin{figure}[h]
\includegraphics[scale= 1]{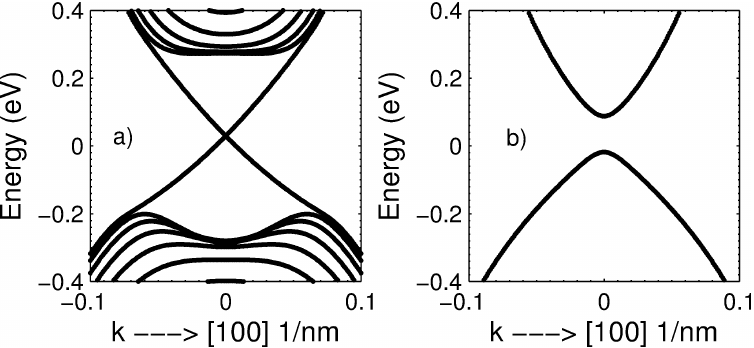}
\caption{Topological insulator surface states (~\ref{fig4}a) around 0.02 $\mathrm{eV}$ for a 20.0 $\mathrm{nm}$ thick (around 20 quintuple layers) Bi$_{2}$Se$_{3}$ film. The dispersion of the thin film (Fig.~\ref{fig4}b) shows two Dirac hyperbolas when the surface states hybridize.}
\label{fig4}
\end{figure}

 The dispersion for a 20.0 $\mathrm{nm}$ Bi$_{2}$Se$_{3}$ thick film which is approximately twenty quintuple-layers is shown in Fig.~\ref{fig4}a. The Dirac cone is formed at an energy equal to 0.029 $\mathrm{eV}$ confirming that it is indeed a mid-gap state. The bulk band-gap of Bi$_{2}$Se$_{3}$ is approximately 0.32 $\mathrm{eV}$ at the $\Gamma$ point. In contrast to the thick-film dispersion, the band profile of a 3.0 $\mathrm{nm}$ (approximately three quintuple-layers) Bi$_{2}$Se$_{3}$ film has a finite band gap. In the case of a thin-film, the two surface states hybridize. The hybridization occurs because each state has a definite localization or penetration length. When the penetration length is comparable to film thickness, the opposite spin-resolved bands of the two surfaces will mix. Since bands with identical quantum numbers cannot cross, a gap(Fig.~\ref{fig4}b) opens up at the $\Gamma$ point~\cite{lu2011weak,li2010chern} and the dispersion changes to Dirac-hyperbolas. A hallmark of these surface states is the inherent spin-momentum locking. The spin-polarization has no out-of-plane spin component~\cite{yazyev2010spin,hsieh2009tunable} and is completely contained in the two-dimensional plane. Spin-polarization calculations starting from a four-band k.p Hamiltonian are shown elsewhere.~\cite{sengupta2014proximity} 
  
\section{Conclusions} 
In this work, a finite difference technique to discretize a continuum k.p Hamiltonian is described. This method is demonstrated by considering a quantum dot heterostructure and topological insulator slab. k.p calculations, specially for significantly large systems enjoy considerable leverage in terms of compute time. For the quantum dot heterostructure, the current k.p simulation with a 1.0 $\mathrm{nm} $ homogeneous spacing resulted in a total of 216,000 nodes. A corresponding twenty-band sp$^{3}$d$^{5}$s$^{*}$ simulation is performed on a subset of nine million atomic positions. As a result calculations are roughly six times faster in the continuum case over their atomistic counterpart. The advantage in terms of computational resources with a continuum k.p calculations is certainly impressive but it suffers from an incomplete description. The k.p Hamiltonian can only describe the energy dispersion around a high-symmetry point and not the complete Brillouin zone. Additionally, the symmetry of the underlying crystal is adequately represented. For instance, zincblende crystal which belongs to the $ C_{2v} $ point-symmetry group is represented as $ C_{4v} $ which possesses higher symmetry. A more detailed description for zincblende quantum dot heterostructures and topological insulators is given elsewhere.

The presented results were obtained with an extended version of NEMO-3D-Peta.~\cite{ryu2009study} To reduce compute times by utilization of hundreds to thousands of computing cores in parallel, a 3D-domain decomposition scheme is employed. A device can be spatially decomposed into three dimensions and each sub-domain is assigned to a corresponding CPU. Based on the spatial information, each CPU only stores the information of the atoms in its sub-domain and neighbouring atoms from adjacent sub-domains; no global position information is held locally, minimizing memory consumption and enabling the simulation of large devices. The reader is referred elsewhere for a more complete discussion.~\cite{lee2011electronic,lee2009million}

\begin{acknowledgements}
Computational resources from nanoHUB.org and support by National Science Foundation (NSF) (Grant Nos. EEC-0228390, OCI-0749140) are acknowledged. This work was also supported by the Semiconductor Research Corporation's (SRC) Nanoelectronics Research Initiative and National Institute of Standards \& Technology through the Midwest Institute for Nanoelectronics Discovery (MIND), SRC Task 2141, and Intel Corporation. One of us (PS) wishes to thank Muhammad Usman for sharing his tight-binding data on quantum dots.
\end{acknowledgements}

\bibliographystyle{apsrev}

\begin{appendices}
\appendix
\renewcommand \thesubsection{\Roman{subsection}}
\titlespacing\section{5pt}{12pt plus 4pt minus 2pt}{0pt plus 2pt minus 2pt}
\section{Basis states}
The angular momentum $\vert J,M_{J}\rangle $ dependent basis states for conduction band, heavy hole, light hole, and spin split-off bands~\cite{winkler2003spin} used for creating the eight-band Hamiltonian (Eq(~\ref{zbham})) for zinc blende crystals is given below. The axis of quantization of angular momentum is assumed to lie along the \textit{z}-axis. Combining \textit{p}-like orbitals ($ l = 1 $) with spinors ($ s = \dfrac{1}{2}$) gives rise to four-fold degenerate $ j = \dfrac{3}{2} $ or two-fold degenerate $ j = \dfrac{1}{2} $ states.
\begin{gather*}
\vert 1\rangle = \vert iS\uparrow\rangle \\
\vert 2\rangle = \dfrac{-1}{\sqrt{2}} \vert \left( X + iY\right) \uparrow\rangle = \vert \dfrac{3}{2},\dfrac{3}{2}\rangle \\
\vert 3\rangle = \dfrac{1}{\sqrt{6}} \vert -\left( X + iY\right) \downarrow + 2Z \uparrow \rangle = \vert \dfrac{3}{2},\dfrac{1}{2}\rangle \\
\vert 4\rangle = \dfrac{1}{\sqrt{3}} \vert \left( X + iY\right) \downarrow + Z \uparrow \rangle = \vert \dfrac{1}{2},\dfrac{1}{2}\rangle \\
\vert 5\rangle = \vert iS\downarrow\rangle \\
\vert 6\rangle = \dfrac{1}{\sqrt{2}} \vert \left( X - iY\right) \downarrow\rangle = \vert \dfrac{3}{2},-\dfrac{3}{2}\rangle \\
\vert 7\rangle = \dfrac{1}{\sqrt{6}} \vert \left( X - iY\right) \downarrow + 2Z \downarrow \rangle = \vert \dfrac{3}{2},-\dfrac{1}{2}\rangle \\
\vert 8\rangle = \dfrac{1}{\sqrt{3}} \vert \left( X - iY\right) \downarrow - Z \downarrow \rangle = \vert \dfrac{1}{2},-\dfrac{1}{2}\rangle  \tag{A1} 
\label{zbbasis}
\end{gather*}

\section{Coefficient matrices}
In this appendix, coefficient matrices defined in Eq(~\ref{coeff_matrix}) are explicitly shown. For sake of brevity, representative matrices for $ k_{x}^{2} $, $ k_{x} $, and $ k_{x}k_{y} $ are derived. The other matrices can be similarly constructed. Note that the coefficient matrix of $ k_{x}^{2} $ and $ k_{y}^{2} $ are identical. Further, coefficient matrices of $ k_{x}^{2}, k_{y}^{2} $ and $ k_{z}^{2} $ are diagonal. The $ Const $ matrix for wire in Eq(~\ref{coeff_matrixwire}) is also included. 

The k.p Hamiltonian can be conveniently turned into a single-band effective mass Hamiltonian if coefficient matrices $C_{i} $ for terms $ k_{i}k_{_{j}} $ are set to zero. These matrices have been identified within the text as $ M_{p,q,r}^{'}$. Cross terms in the k.p Hamiltonian lead to warping~\cite{altarelli1985calculations} of valence bands in a realistic way and cannot be captured by an effective mass Hamiltonian. All symbols used here have the identical meaning as defined in Eq(~\ref{symb}).

\subsection{Coefficient of $ k_{x}^{2} $ }
\begin{widetext}
\[
A_{1 (8 \times 8)} =  \left( \begin{array}{cccccccc}
t_{1} & 0 & 0 & 0 & 0 & 0 & 0 & 0 \\
0 & -t_{0}(\gamma_{1}+\gamma_{2}) & 0 & 0 & 0 & 0 & 0 & 0 \\
0 & 0 & t_{0}(- \gamma_{1}+\gamma_{2}) & 0 & 0 & 0 & 0 & 0 \\
0 & 0 & 0 & -t_{0}\gamma_{1} & 0 & 0 & 0 & 0 \\
0 & 0 & 0 & 0 & t_{1} & 0 & 0 & 0 \\
0 & 0 & 0 & 0 & 0 & t_{0}(- \gamma_{1}+\gamma_{2}) & 0 & 0\\ 
0 & 0 & 0 & 0 & 0 & 0 & t_{0}(\gamma_{1}+\gamma_{2}) & 0 \\
0 & 0 & 0 & 0 & 0 & 0 & 0 &  -t_{0}\gamma_{1} \tag{B1}
\end{array} \right)
\label{a1}
\]
\end{widetext}
where
\begin{gather} 
t_{1} = \left(\dfrac{\hbar^{2}}{2m^{*}_{e}}\right) \notag ;
t_{0} = \left(\dfrac{\hbar^{2}}{2m^{*}_{0}}\right) \notag
\end{gather}
\subsection{Coefficient of $ k_{x} $ }
\begin{widetext}
\[
B_{1 (8 \times 8)} =  \left( \begin{array}{cccccccc}
0 & -\dfrac{P_{cv}}{\sqrt{2}} & 0 & 0 & 0 & 0 & \dfrac{P_{cv}}{\sqrt{6}} & \dfrac{P_{cv}}{\sqrt{3}} \\
0 & 0 & 0 & 0 & 0 & 0 & 0 & 0 \\
0 & 0 & 0 & 0 & -\dfrac{P_{cv}}{\sqrt{6}} & 0 & 0 & 0 \\
0 & 0 & 0 & 0 & \dfrac{P_{cv}}{\sqrt{3}} & 0 & 0 & 0 \\
0 & 0 & -\dfrac{P_{cv}}{\sqrt{6}} & \dfrac{P_{cv}}{\sqrt{3}} & 0 & 0 & 0 & 0 \\
0 & 0 & 0 & 0 & \dfrac{P_{cv}}{\sqrt{2}}  & 0 & 0 & 0\\ 
\dfrac{P_{cv}}{\sqrt{6}} & 0 & 0 & 0 & 0 & 0 & 0 & 0 \\
\dfrac{P_{cv}}{\sqrt{3}} & 0 & 0 & 0 & 0 & 0 & 0 & 0 \tag{B2}
\end{array} \right)
\label{b1}
\]
\end{widetext}

\subsection{Coefficient of $ k_{x}k_{y} $ }
\begin{widetext}
\[
C_{1 (8 \times 8)} =  \left( \begin{array}{cccccccc}
0 & 0 & 0 & 0 & 0 & 0 & 0 & 0 \\
0 & 0 & 0 & 0 & 0 & 0 & -it_{0}2\sqrt{3}\gamma_{3} & -it_{0}\sqrt{24}\gamma_{3}  \\
0 & 0 & 0 & 0 & 0 &  -it_{0}2\sqrt{3}\gamma_{3} & 0 & 0 \\
0 & 0 & 0 & 0 & 0 & it_{0}\sqrt{24}\gamma_{3} & 0 & 0 \\
0 & 0 & 0 & 0 & 0 & 0 & 0 & 0 \\
0 & 0 & it_{0}2\sqrt{3}\gamma_{3}  & -it_{0}\sqrt{24}\gamma_{3} & 0 & 0 & 0 & 0\\ 
0 & it_{0}2\sqrt{3}\gamma_{3} & 0 & 0 & 0 & 0 & 0 & 0 \\
0 & -it_{0}\sqrt{24}\gamma_{3} & 0 & 0 & 0 & 0 & 0 & 0 \tag{B3}
\end{array} \right)
\label{c1}
\]
\end{widetext}
\subsection{Const matrix for a wire}
\begin{widetext}
\begin{equation}
Const =  \begin{pmatrix} 
E_{c}+t_{1}k_{z}^{2} & 0 & \sqrt{2}U & U & 0 & 0 & 0 & 0 \\
0 & \lambda_{1} & 0 & 0 & 0 & 0 & 0 & 0 \\
\sqrt{2}U & 0 & \lambda_{2} & -2\sqrt{2}t_{0}\gamma_{2}k_{z}^{2} & 0 & 0 & 0 & 0 \\
U & 0 & -2\sqrt{2}t_{0}\gamma_{2}k_{z}^{2} &  \lambda_{3} & 0 & 0 & 0 & 0 \\
0 & 0 & 0 & 0 & E_{c}+t_{1}k_{z}^{2} & 0 &  \sqrt{2}U & -U \\
0 & 0 & 0 & 0 & 0 &  \lambda_{1} & 0 & 0 \\ 
0 & 0 & 0 & 0 & \sqrt{2}U & 0 & \lambda_{2} & -\sqrt{2}Q \\
0 & 0 & 0 & 0 & -U & 0 & -\sqrt{2}Q &  \lambda_{3} \tag{B4}
\label{constwire}
\end{pmatrix}
\end{equation}
\end{widetext} 
where
\begin{gather*} 
\lambda_{1} = E_{v}-t_{0}(\gamma_{1}+2\gamma_{2})k_{z}^{2} \notag \\
\lambda_{2} = E_{v}-t_{0}(\gamma_{1}-2\gamma_{2})k_{z}^{2}  \notag \\
\lambda_{3} = E_{v}-t_{0}\gamma_{1}k_{z}^{2} -\Delta \tag{B5}
\end{gather*}
\end{appendices}

\end{document}